\documentclass{cl2emult}

\usepackage{makeidx}  
\usepackage{graphicx} 
\usepackage{subeqnar} 
\usepackage{multicol} 
\usepackage{cropmark} 
\usepackage{eso}      
\makeindex            

\begin{document}

\def\omegam{\Omega_{\rm Macho}}
\def\omegab{\Omega_{\rm B}}
\def\lya{Ly$\alpha$}


\title*{\bf Death of Stellar Baryonic Dark Matter}

\author{Katherine Freese\inst{1}
\and Brian Fields\inst{2}
\and David Graff\inst{3}}
\authorrunning{Katherine Freese et al.}
\institute{University of Michigan, Dept. of Physics, Ann Arbor, MI 48109-1120
\and Ohio State University, Astronomy Dept., Columbus, OH   43210
\and University of Illinois, Astronomy Dept., Urbana, IL   61801-3080}
\maketitle
\begin{abstract}

The nature of the dark matter in the haloes of galaxies is one of the
outstanding questions in astrophysics.  All stellar candidates, until
recently thought to be likely baryonic contributions to the Halo of
our Galaxy, are shown to be ruled out.  Faint stars and brown dwarfs
are found to constitute only a few percent of the mass of the Galaxy.
Stellar remnants, including white dwarfs and neutron stars, are shown
to be very constrained as well.  High energy gamma-rays observed in
HEGRA data place the strongest constraints, $\Omega_{WD} < 3 \times
10^{-3} h^{-1}$, where $h$ is the Hubble constant in units of 100 km
s$^{-1}$ Mpc$^{-1}$.  Hence one is left with several unanswered
questions: 1) What are MACHOs seen in microlensing surveys? 2) What is
the dark matter in our Galaxy?  Indeed a nonbaryonic component in the
Halo seems to be required.

\end{abstract}

\section{Introduction}

The nature of the dark matter in the haloes of galaxies is an
outstanding problem in astrophysics.  Over the last several decades
there has been great debate about whether this matter
is baryonic or must be exotic.  Many astronomers believed
that a stellar or substellar solution to this problem might be
the most simple and therefore most plausible explanation.
However, in the last few years, these candidates have been
ruled out as significant components of the Galactic Halo.
I will discuss limits on these stellar candidates, and
argue for my personal conviction that:
{\em Most of the dark matter in the Galactic Halo must be nonbaryonic.}

Until recently, stellar candidates for the dark matter, including
faint stars, brown dwarfs, white dwarfs, and neutron stars, were
extremely popular.  However, recent analysis of various data sets has
shown that faint stars and brown dwarfs probably constitute no more
than a few percent of the mass of our Galaxy
\cite{kfrbfgk,kfrgf96a,kfrgf96b,kfrmcs,kfrfgb,kfrfreese}.
Specifically, using Hubble Space Telescope and 
parallax data, we showed
that faint stars and brown dwarfs contribute no more than 1\% of the
mass density of the Galaxy.  Microlensing experiments
(the MACHO \cite{kfrmacho:1yr}, \cite{kfrmacho:2yr} and EROS
\cite{kfransari}) experiments), which were
designed to look for Massive Compact Halo Objects (MACHOs), also
failed to find these light stellar objects and ruled out
substellar dark matter candidates in the $(10^{-7} - 10^{-2}) M_\odot$ mass
range.

Recently white dwarfs have received attention as possible dark matter
candidates.  Interest in white dwarfs has been motivated by
microlensing events interpreted as being in the Halo, with a best fit
mass of $\sim 0.5 M_\odot $.  
However, I will show that stellar remnants including white dwarfs
and neutron stars are extremely problematic as dark matter
candidates. A combination of excessive infrared radiation,
mass budget issues and chemical abundances constrains the
abundance of stellar remnants in the Halo quite severely,
as shown below.   
Hence, white dwarfs, brown dwarfs, faint stars, and neutron stars are either ruled
out or extremely problematic as dark matter candidates.  Thus the
puzzle remains, What are the 14 MACHO events that have been
interpreted as being in the Halo of the Galaxy?  Are some of them
actually located elsewhere, such as in the LMC itself? These questions
are currently unanswered.  As regards the dark matter in the Halo of
our Galaxy, one is driven to nonbaryonic constituents as the bulk of
the matter. Possibilities include supersymmetric particles, axions,
primordial black holes, or other exotic candidates.

In this talk I will focus on the arguments against stellar remnants
as candidates for a substantial fraction of the dark matter,
as white dwarfs in particular have been the focus of attention
as potential explanations of microlensing data.  For a discussion
of limits on faint stars and brown dwarfs, see earlier
conference proceedings by Freese, Fields, and Graff (\cite{kfrfreese} and
\cite{kfrconf1}).

\section{White Dwarfs}

Stellar remnants (white dwarfs and neutron stars) face
a number of problems and issues as dark matter candidates:
1) infrared radiation;
2) IMF (initial mass function);
3) baryonic mass budget; 
4) element abundances. 

We find that none of the expected signatures in the above
list of a significant white dwarf component in the Galactic
Halo are seen to exist.

\subsection{Constraints from multi-TeV $\gamma$-rays seen by HEGRA}

The mere existence of multi-TeV $\gamma$-rays seen in
the HEGRA experiment places a powerful constraint on the
allowed abundance of white dwarfs.  This arises because
the progenitors of the white dwarfs would produce infrared
radiation that would prevent the $\gamma$-rays from getting here.
The $\gamma$-rays and infrared photons would interact via
$\gamma \gamma \rightarrow e^+ e^-$.  

Multi-TeV $\gamma$-rays from the blazar Mkn 501 at a redshift z=0.034
are seen in the HEGRA detector.  The
cross section for (1-10)TeV $\gamma$-rays peaks at infrared photon
energies of (0.03-3)eV.  Photons in this energy range would be
produced in abundance by the progenitor stars to white dwarfs and
neutron stars.  
By requiring that the optical depth due to
$\gamma \gamma \rightarrow e^+ e^-$ be less than one for a source at
$z=0.034$ we limit the cosmological density of stellar remnants
\cite{kfrgfwp} to
\(\Omega_{\rm WD} \leq (1-3) \times 10^{-3}
h^{-1}\).
This constraint is quite robust and model independent, as it applies
to a variety of models for stellar physics, star formation rate and
redshift, mass function, and clustering.

\subsection{Mass Budget Issues}

\paragraph{Contribution of Machos to the Mass Density of the
Universe:}
(based on work by Fields, Freese, and Graff \cite{kfrffg})
There is a potential problem in that too many baryons are
tied up in Machos and their progenitors (Fields, Freese, and Graff).
We begin by estimating the contribution of Machos to the mass density of the
universe.
Microlensing results \cite{kfrmacho:1yr} predict that the total mass
of Machos in the Galactic Halo out to 50 kpc is
$M_{\rm Macho} = (1.3 - 3.2) \times 10^{11} M_\odot  \, .$
Now one can obtain a ``Macho-to-light" ratio for the Halo by
dividing by the luminosity of the Milky Way (in the B-band),
$L_{MW} \sim (1.3-2.5) \times 10^{10} L_\odot,$
to  obtain
$(M/L)_{\rm Macho} = (5.2-25)M_\odot /L_\odot \, .$
>From the ESO Slice Project
Redshift survey \cite{kfrzuc},
the luminosity
density of the Universe in the $B$ band is
${\cal L}_B = 1.9\times 10^{8} h \ L_\odot \ {\rm Mpc}^{-3} \, .$
If we assume that the $M/L$ which we defined for the Milky
Way is typical of the Universe as a whole,
then the universal mass density of
Machos is
\begin{equation}
\label{kfromega}
\Omega_{\rm Macho} \equiv \rho_{\rm Macho}/ \rho_c = (0.0036-0.017) \,
h^{-1} \, 
\end{equation}
where the critical density
$\rho_c \equiv
3H_0^2/8 \pi G = 2.71 \times 10^{11} \, h^2 \, M_\odot \ {\rm Mpc}^{-3}$.

We will now proceed to compare our $\omegam$
derived in Eq.~\ref{kfromega} with the baryonic density in the universe,
$\omegab$, as determined by primordial nucleosynthesis.
To conservatively allow for the full range of possibilities,
we will adopt
$\omegab= (0.005-0.022) \ h^{-2} \, .$
Thus, if the Galactic halo Macho
interpretation of the microlensing
results is correct,
Machos make up an important fraction of the baryonic matter
of the Universe.
Specifically, the central values give
\begin{equation}
\omegam/\omegab \sim 0.7\, .
\end{equation}
However, the lower limit on this fraction is
considerably less restrictive,
\begin{equation}
{\omegam \over \omegab} \geq {1 \over 6} h \geq \frac{1}{12}\, .
\end{equation}

\paragraph{Mass Budget constraints from
Machos as Stellar Remnants: White Dwarfs or Neutron Stars}

In general, white dwarfs, neutron stars, or black holes all came from
significantly heavier progenitors.  Hence, the excess mass left over
from the progenitors must be added to the calculation of $\Omega_{\rm
Macho}$; the excess mass then leads to stronger constraints.
Typically we find the contribution of Macho progenitors to the mass
density of the universe to be
\( \Omega_{{\rm prog}} = 4 \Omega_{{\rm Macho}} = (0.016-0.08)h^{-1}\).
The central values of all the numbers now imply
$\Omega_{\rm prog} \sim 3 \Omega_B$, 
which is obviously unacceptable.  One is driven to the lowest
values of $\Omega_{\rm Macho}$ and highest value of $\Omega_B$
to avoid this problem.

\subsection{On Carbon and Nitrogen}

The overproduction of carbon and/or nitrogen
produced by white dwarf progenitors is one of the
greatest difficulties faced by a white dwarf dark matter scenario,
as first noted by Gibson and Mould \cite{kfrgm}.
Stellar carbon yields for zero
metallicity stars are quite uncertain.
Still, according to the yields by \cite{kfrvdhg}, a star of mass
2.5$M_\odot $ will produce about twice the
solar enrichment of carbon.
However,  stars in our galactic halo have carbon
abundance in the range $10^{-4}-10^{-2}$ solar.
Hence the ejecta of a
large population of white dwarfs would have to be removed
from the galaxy via a galactic wind.

However, carbon abundances in intermediate redshift
\lya\ forest lines have recently been measured to be
quite low, at the
$\sim 10^{-2}$ solar level
\cite{kfrsc}, for \lya\ systems at $z \sim 3$
with column densities $N \ge 3 \times 10^{15} \, {\rm cm}^{-2}$
(for lower column densities, the mean C/H drops to $\sim 10^{-3.5}$ solar
\cite{kfrlsbr}.

\begin{figure}[ht]
\centering
\includegraphics[scale=0.3,bb= 0 80 590 720]{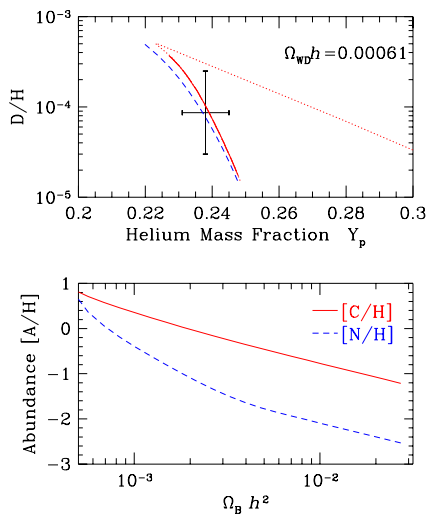}
\caption{{\small (taken from Fields, Freese, and Graff 1999): {\bf
(a)} The D/H abundances and helium mass fraction $Y$ for models with
$\Omega_{\rm WD} h = 6.1 \times 10^{-4}$, $h=0.7$, and IMF peaked at
$2M_\odot $.  
The short-dashed curve shows the initial abundances and the error bars
the range of D and He measurements. 
The other three curves show the changes in primordial D and He as a
result of white dwarf production. The solid one is for the full
chemical evolution model, the dotted one for instantaneous
recycling, and the long-dashed one for the burst model.   
This is the absolute minimum
$\Omega_{\rm WD}$ compatible with cosmic extrapolation of white dwarf
Machos if Machos are contained only in spiral galaxies with luminosities
similar to the Milky Way.  {\bf (b)} CNO abundances
produced in the same model as {\bf a}, here plotted as a function of
$\Omega_B$.  The CN abundances are
presented relative to solar via the usual notation of the form
$[{\rm C/H}]= \log_{10} \frac{{\rm C/H}}{({\rm C/H})_\odot} \, .$
The C and N production in particular are greater than 1/10
solar.}}
\end{figure}

In order to maintain carbon abundances as low as $10^{-2}$ solar, only
about $10^{-2}$ of all baryons can have passed through the
intermediate mass stars that were the predecessors of Machos
\cite{kfrffg}.  Such a
fraction can barely be accommodated for
the remnant density predicted from our extrapolation of the Macho
group results, and would be in conflict with $\Omega_{{\rm prog}}$ in
the case of a single burst of star formation.  Note that
stars heavier than 4$M_\odot $ may replace the carbon overproduction
problem with nitrogen overproduction
\cite{kfrvdhg,kfrlattbooth}.

Using the yields described above, we calculated the C and N that would
result from the stellar processing for a variety of initial mass
functions for the white dwarf progenitors.  We used a chemical
evolution model based on a code described in Fields \& Olive
\cite{kfrfo98} to obtain our numerical results.  Our results are
presented in the figure.

In the figure, we make the parameter choices that are in agreement with D
and He$^4$ measurements (see the discussion below) and are the least
restrictive when comparing with the Ly$\alpha$ measurements.
We take an initial mass function (IMF)
sharply peaked at 2$M_\odot $, so that there are very few progenitor
stars heavier than 3$M_\odot $ (this IMF is required by D and He$^4$
measurements).  In addition (see the figures in Fields,
Freese, and Graff \cite{kfrffg2}) we have considered a variety of other
parameter choices.  By comparing with the observations, we obtain the limit,
\(\Omega_{\rm WD} h \leq 2 \times 10^{-4}\).
As a caveat, note that it is possible that carbon never leaves the
(zero metallicity)
white dwarf progenitors, so that carbon overproduction is not a
problem \cite{kfrchabriernew}.  

\subsection{Deuterium and Helium}

Because of the uncertainty in the C and N yields from low-metallicity
stars, we have also calculated the D and He$^4$ abundances that would
be produced by white dwarf progenitors.  These are far less uncertain
as they are produced farther out from the center of the star and do
not have to be dredged up from the core.  
Panel a) in the figure displays our results.  Also shown are the initial
values from big bang nucleosynthesis and the (very generous) range of
primordial values of D and He$^4$ from observations.
>From D and He alone, we can see that the white dwarf progenitor IMF
must be peaked at low masses, $\sim 2M_\odot $.
We obtain \(\Omega_{\rm WD} \leq 0.003\).



\section{Conclusions}

\paragraph{A Zero Macho Halo?}
The possibility exists that the 14 microlensing events that have been
interpreted as being in the Halo of the Galaxy are in fact due to some
other lensing population.  One of the most difficult aspects of
microlensing is the degeneracy of the interpretation of the data, so
that it is currently impossible to determine whether the lenses lie in
the Galactic Halo, or in the Disk of the Milky Way, or in the LMC.  In
particular, it is possible that the LMC is thicker than previously
thought so that the observed events are due to self-lensing of the
LMC.  All these possibilities are being investigated.  More data
are required in order to identify where the
lenses are.

Microlensing experiments have ruled out baryonic dark matter objects
in the mass range $10^{-7}M_\odot $ all the way up to $10^{-2}M_\odot$.
In this talk
I discussed the heavier possibilities in the range $10^{-2}M_\odot $
to a few $M_\odot $.  Brown dwarfs and faint stars
are ruled out as significant dark matter components; they contribute
no more than 1\% of the Halo mass density.  Stellar remnants
are not able to explain the dark matter of the Galaxy either;
none of the expected
signatures of stellar remnants, i.e., infrared radiation,
large baryonic mass budget, and C,N, and He$^4$ abundances,
are found observationally.

Hence, in conclusion, \hfill\break 1) Nonbaryonic dark matter in our
Galaxy seems to be required, and \hfill\break 2) The nature of the
Machos seen in microlensing experiments and interpreted as the dark
matter in the Halo of our Galaxy remains a mystery.  Are we driven to
primordial black holes \cite{kfrcarr} \cite{kfrjedam},
nonbaryonic Machos (Machismos?), mirror matter Machos (\cite{kfrmohap})
or perhaps a no-Macho Halo?

\end{document}